\documentclass[a4paper]{jpconf}
\usepackage{graphicx}
\begin{document}
\title{GLACIER and related R\&D}

\author{Alessandro Curioni}

\address{ETH Z\"urich}

\ead{alessandro.curioni@cern.ch}

\begin{abstract}

Liquid argon detectors, with mass up to 100 kton, are being actively studied in the context of proton decay searches, neutrino 
astrophysics and for the next generation of long baseline neutrino oscillation experiments to study the neutrino mass hierarchy and CP 
violation in the leptonic sector. The proposed Giant Liquid Argon Charge Imaging ExpeRiment (GLACIER) offers a well defined conceptual 
design for such a detector. In this paper we present the GLACIER design and some of the R\&D activities pursued within the GLACIER.

\end{abstract}

\section{Introduction}

The development of very large liquid argon time projection chambers (LAr TPCs), with a fiducial mass up to 100 kton, is being actively 
pursued in the framework of tackling the following topics: 
\begin{itemize}
\item[a.] neutrino oscillations, in particular to study CP violation in the leptonic sector with a sensitivity far superior to running/approved 
experiments as T2K \cite{Itow:2001ee} and NO$\nu$A \cite{Ayres:2004js};
\item[b.] proton decay searches, where LAr TPCs can provide unprecedented sensitivity especially in the decay channels with a kaon in the 
final state, with good complementarity to searches with water Cherenkov detectors;
\item[c.] astrophysical neutrinos, i.e. neutrinos from stellar collapse, solar neutrinos etc.
\end{itemize}
This is essentially a continuation of the the physics program covered by a large water Cherenkov as Superkamiokande \cite{Wendell:2010md, Thrane:2009tw, Hosaka:2006zd, Hosaka:2005um, Abe:2010hy, Kobayashi:2005pe, Nishino:2008zz, Ashie:2005ik}, which has 
operated successfully for more than a decade, therefore setting strict requirements on the sensitivity of any future experiment.
Liquid argon is a persistent candidate because the LAr TPC technology offers significant improvements on key parameters with respect to 
water Cherenkov detectors, namely: energy resolution, excellent control over background, good signal efficiency over the relevant energy 
range, and a detailed pictorial reconstruction on an event-by-event basis, much as in a bubble chamber.
Another key parameter of LAr TPC is the possibility to scale up to the needed size, of the order of 100 kton in the fiducial volume; for a 
review, see \cite{Curioni:2009zza}. 

So far, the ICARUS \cite{Amerio:2004ze, Arneodo:2006ug, Cennini:1994ha, Benetti:1993yn} experience has produced the largest operating 
LAr TPC, the ICARUS T600, which comprises two separate modules for a total mass of about 600 ton and is now taking data in the LNGS.
In order to deliver the needed sensitivity, a next generation LAr TPC has to be scaled up by a factor larger than 100 from the current state of
the art. Various proposals exist to bridge the gap in detector mass; here we present the GLACIER concept and related R\&D activities.

\section{GLACIER}

The basic concept of GLACIER was originally put forward in \cite{Rubbia:2004tz}, where many details can be found.
The key points of the proposed design are:
\begin{itemize}
\item single module non-evacuable cryogenic tank based on industrial liquefied natural gas (LNG) technology, of cylindrical shape with 
excellent surface/volume ratio;
\item simple, scalable detector design, possibly up to 100 kton; 
\item single very long vertical drift (up to 20 m) with full active mass; 
\item for the readout structure, a very large area, up to 3500 m$^2$, instrumented with Large Electron Multipliers (LEM) operating in double 
phase argon (liquid-vapor);
\item possibly immersed visible light readout for Cherenkov imaging;
\item possibly immersed (high T$_c$) superconducting solenoid to obtain a magnetized detector; 
\item reasonable excavation requirements ($<$250,000 m$^3$).
\end{itemize}

Such a design requires a suite of R\&D activities to asses the feasibility of the concept. These activities are the topic of this work, and four of 
them are presented in more detail:
\begin{itemize}
\item[1.] LEM R\&D work at CERN;
\item[2.] the ArDM Dark Matter experiment;
\item[3.] experimental study of Ar purity in a non-evacuable 6m$^3$ vessel;
\item[4.] the T32@J-PARC experiment.
\end{itemize}

\subsection{LAGUNA}
The critically important item of a proper underground location has been extensively addressed in the framework of LAGUNA (Large 
Apparatus for Grand Unification and Neutrino Astrophysics) \cite{Autiero:2007zj, Rubbia:2010zz, Kisiel:2009zz, Angus:2010sz}. The 
LAGUNA design study is dedicated to the feasibility of very large underground infrastructures able to host the next generation neutrino 
physics, astroparticle physics and proton decay experiments. It is conducted in the spirit of an European-wide coordination and is supported 
by the EC FP7 program \footnote{http://www.laguna-science.eu/}. 
The LAGUNA design study considers three different detector technologies (liquid scintillator, liquid argon and water Cherenkov), and seven 
potential underground sites, in order to identify the scientifically and technically most appropriate and cost-effective strategy for future large 
scale underground detectors.
Seven technical reports have been produced and made available. 

\section{LEM R\&D}

In a LAr TPC, the charge produced by ionizing particles in liquid argon is drifted by an applied electric field towards the 
readout electrodes. In the case of a  LAr LEM-TPC the readout electrodes are in the vapor phase, above the liquid surface. 
Therefore the electrons are extracted at the liquid-vapor interface by means of an appropriate electric field. In the 
vapor phase, Townsend avalanche takes place in the high electric field regions confined in the LEM holes, similar to the situation of the Gas 
Electron Multiplier (GEM) \cite{Sauli:1997qp}. The LEM is a macroscopic hole electron multiplier built with standard PCB techniques. The 
amplified charge induces a detectable signal on a set of segmented electrodes, proportional in amplitude to the ionization charge, therefore 
giving both calorimetric and spatial information. 

\begin{figure}[h]
\includegraphics[width=30pc]{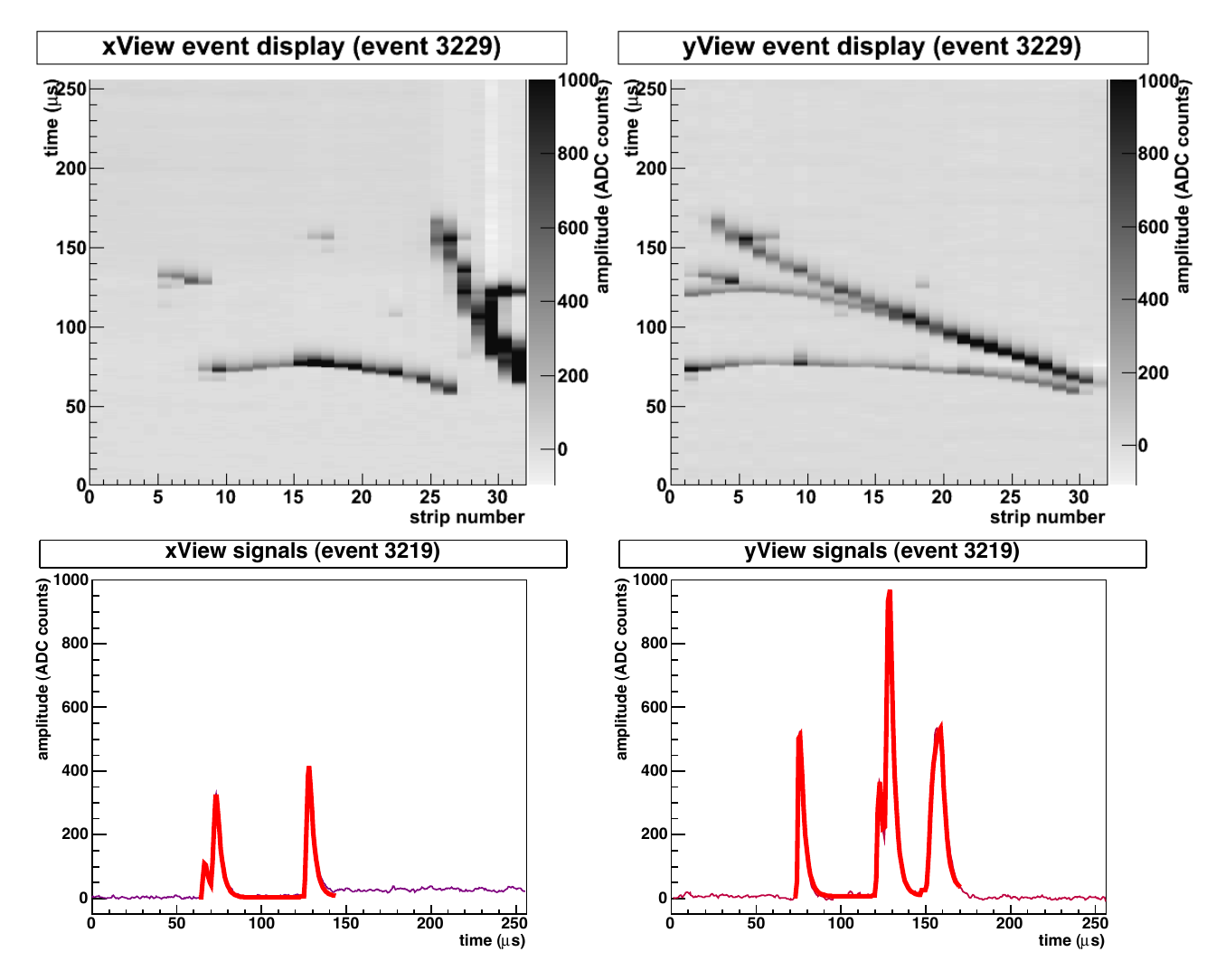}
\caption{\label{f_evt} Multi-track event. Top: Event display: channel number vs. drift time, X-view left, and Y-view right. The gray scale is proportional to the signal amplitude. Bottom: Typical waveforms for the X-view left and the Y-view right. The thick curve shows the result of the signal fit using the response of the preamplifier. From \cite{Badertscher:2010zg}.}
\end{figure}

%
A 3L LAr LEM-TPC has been built and operated at CERN to test different LEM designs under realistic conditions. 
In this setup, the LAr drift volume has a cross-section of 10$\times$10 cm$^2$ and the distance between cathode and LAr surface can be 
extended up to 30 cm. The LAr scintillation light is detected by a Hamamatsu R6237-01 photomultiplier tube, coated with the wavelength 
shifter tetraphenylbutadiene (TPB) and placed below the optically transparent cathode grid, electrically shielded by a grounded protection 
grid. The uniform drift field is defined by equally-spaced rectangular field shaping rings, one every 5 mm, connected to a chain of 940 
M$\Omega$ resistors. On top of the drift volume there are two parallel extraction grids with a wire pitch of 5 mm and a gap of 10 mm 
between them, the lower one immersed in LAr and the  upper one in Ar vapor, which allow to apply an electric field across the LAr surface, 
and to set it independently of the drift field. The level of the LAr is controlled with millimeter precision by a capacity measurement of the two 
grids. The LEM readout is mounted 10 mm above the upper grid. 
After testing several LEM configurations \cite{Badertscher:2010fi, Badertscher:2009av, Badertscher:2008rf} , a very promising design has 
been successfully selected, with a single 1 mm thick LEM amplifying stage and a two dimensional projective readout anode with 3 mm pitch 
readout \cite{Badertscher:2010zg}. A stable gain larger than 25 has been achieved with this detector configuration corresponding to a 
signal-to-noise ratio exceeding 200 for minimum ionizing tracks. An example of an event recorded in this configuration is shown in Fig.~
\ref{f_evt}. It should be stressed that with the 2D projective anode the two orthogonal views are completely equivalent, while in a LAr TPC 
with a wire chamber readout one typically has to distinguish between the collection view and the induction view, with different behavior in 
terms of signal shape and signal-to-noise ratio.

\begin{figure}[h]
\includegraphics[width=20pc]{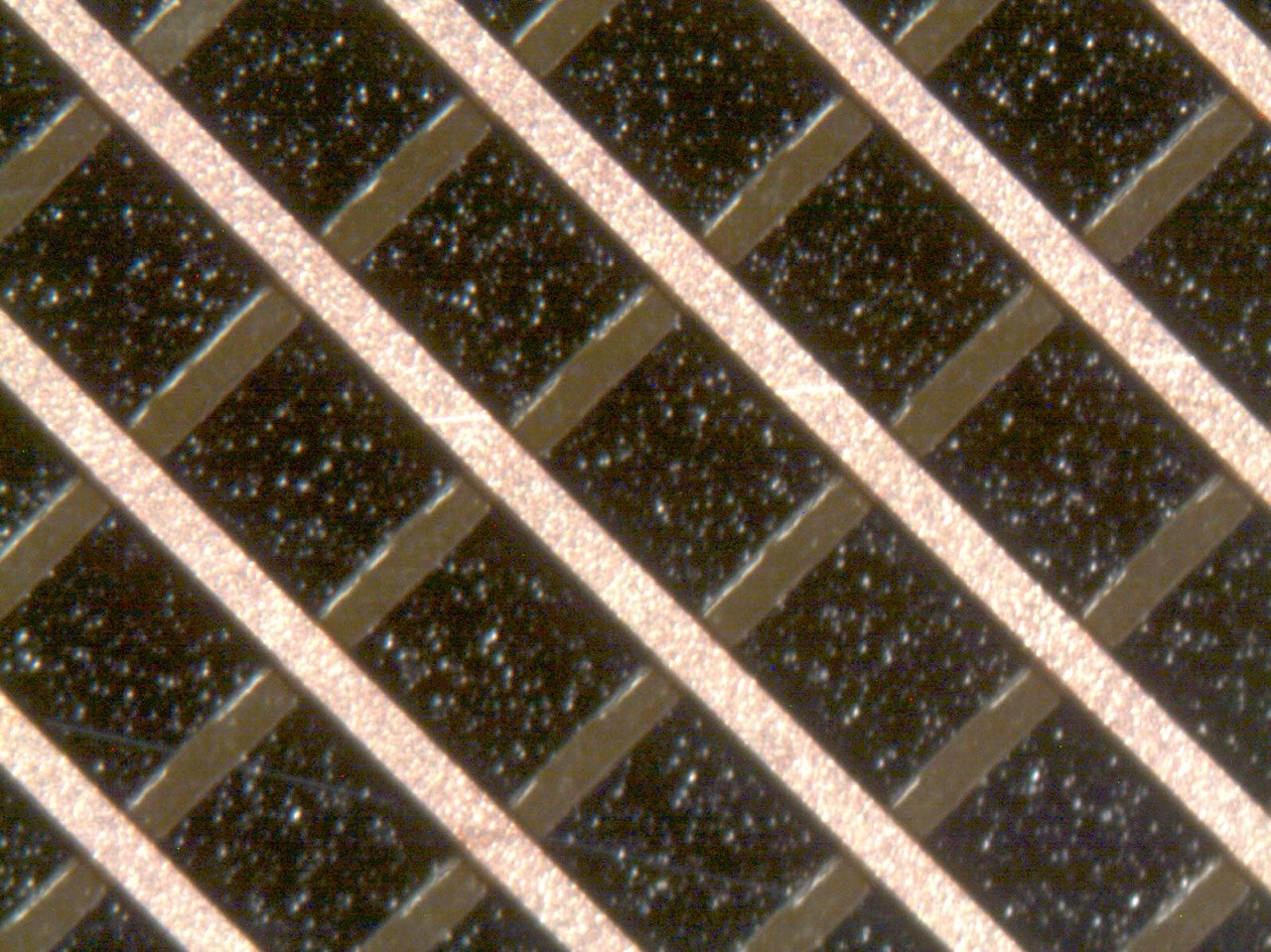}
\includegraphics[width=22pc]{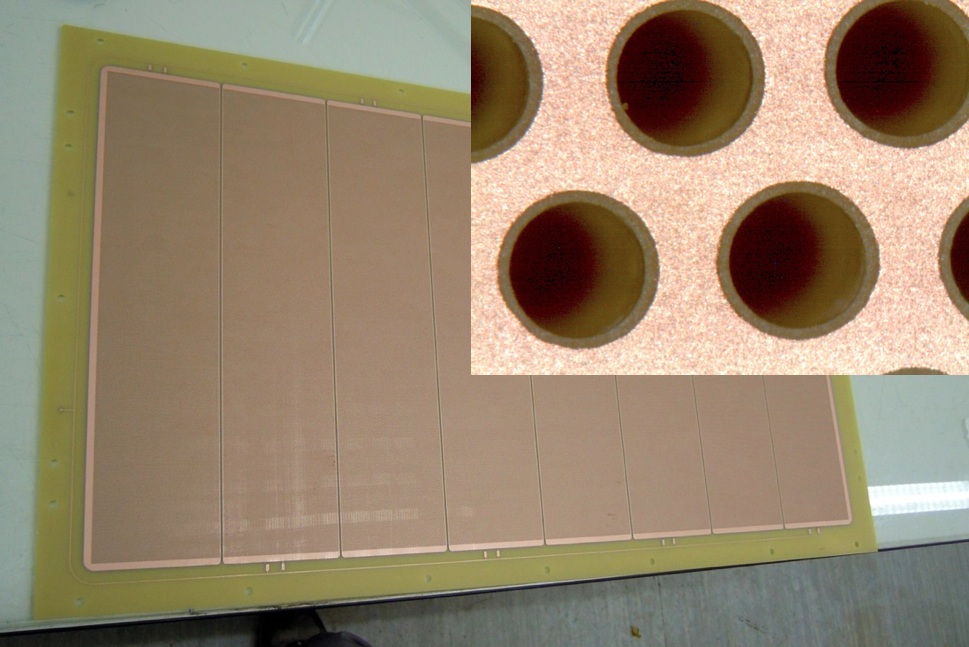}
\caption{\label{f_anode} Left: Close up of the 40$\times$80 cm$^2$ 2D projective anode. Right: 40$\times$80 cm$^2$  LEM, and close up of the holes. The well centered rims around the holes are visible.}
\end{figure}

For a GLACIER type detector, square-meter size LEM and projective anode are needed, in order to cover the entire readout plane with a 
large but still acceptable number of modules. A 40$\times$80 cm$^2$ "readout sandwich", i.e. LEM plus matching 2D projective anode (see 
Fig.~\ref{f_anode}), has been produced, with the same geometry of the 10$\times$10 cm$^2$ LEM tested in the 3L LAr LEM-TPC setup and 
described in \cite{Badertscher:2010zg}, i.e.  PCB thickness 1 mm, hole diameter  500 $\mu$m, hole pitch 800 $\mu$m, dielectric rim size $
\sim$50$\mu$m; it is presently being tested at CERN. It will be mounted on the  250L J-Parc@T32 LAr TPC 
(Sec.~6). Square-meter size readout planes are also being built for the ArDM Dark Matter detector (Sec.~4).

\section{ArDM}

ArDM is an experiment designed for the direct search of WIMP Dark Matter \cite{Rubbia:2005ge}. It has been built and commissioned at 
CERN in the past few years and is now approved for  installation in LSC (Laboratorio Subterr\'aneo de Canfranc) in 2011. The ArDM 
detector is a 1-ton \footnote{In the field of direct Dark Matter searches, a 1-ton instrumented mass is a very large one.} two-phase LAr 
detector, with independent charge and light readout optimized for the detection of low energy events, in the 10-100 keV range. 
WIMP interactions on an argon nucleus are expected to produce a nuclear recoil, much similar to the ones from elastic scattering of MeV 
neutrons, with an exponentially falling recoil spectrum. With an energy threshold of 30 keV (for nuclear recoils) and WIMP-nucleus cross 
section of 10$^{-44}$ cm$^2$, not yet excluded by current limits, about one event per day is expected in ArDM.
Together with having a sizeable detector sensitive to low energy interactions, a key parameter for direct Dark Matter searches is also a high 
discrimination power against background in the relevant energy range.
The LEM TPC readout, with mm position resolution on each spatial coordinate, helps background rejection through accurate fiducial volume 
cuts and separate detection of multiple interactions within the same event.
In terms of rate, the background in a carefully designed 1-ton LAr detector will be dominated by the presence of $^{39}$Ar, which is a beta 
emitter (Q-value 565 keV and a half-life of 269 years) present in natural argon with a concentration of 8$\times$10$^{-16}$, resulting in an 
activity of about 1 Bq/kg \cite{Benetti:2006az}. 
LAr offers an extremely good discrimination between beta and nuclear recoils, based on a combined analysis of the charge/light ratio and 
pulse shape for the scintillation light.
In ArDM, the present light readout of the prompt scintillation light is done using 14 cryogenic photomultipliers immersed in LAr. The detailed 
description of this system can be found in \cite{Amsler:2010yp}.

\begin{figure}[h]
\includegraphics[width=30pc]{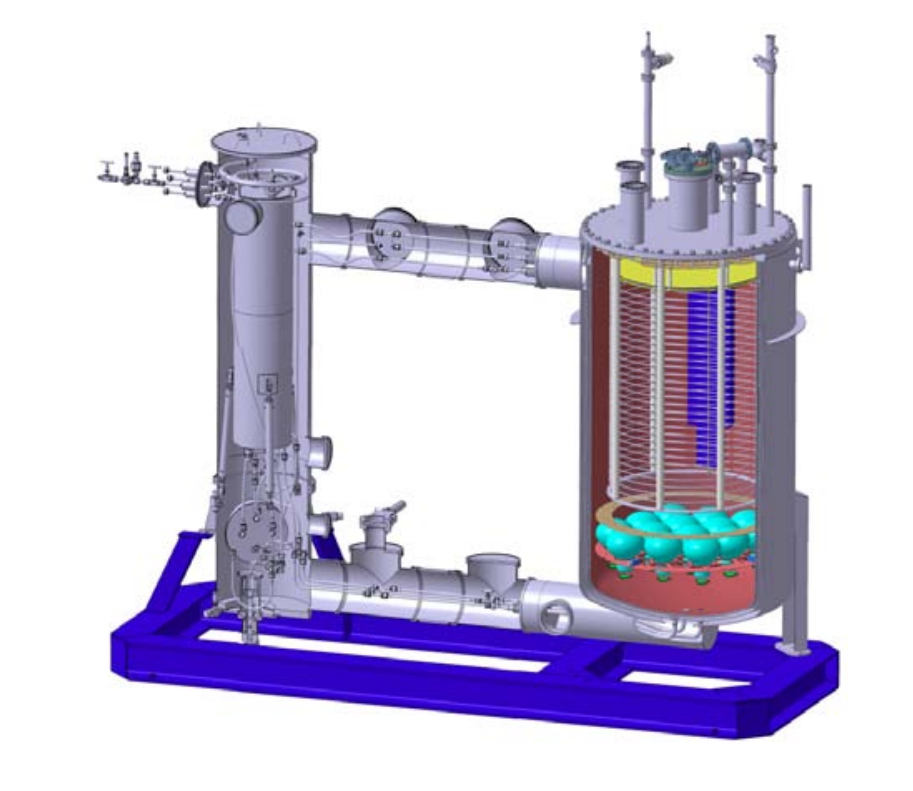}
\caption{\label{f_ardm} 3D of the fully assembled ArDM detector. On the left, the cooling system, on the right, the detector, with the photomultiplier tubes at the bottom of the active volume.}
\end{figure}

While ArDM is a stand-alone experiment pursuing a well defined physics program, some of the developed technical solutions are of 
immediate interest for GLACIER, in particular for the cathode high voltage and the LAr LEM-TPC charge readout. The LEM charge readout 
is very similar to the one envisioned for GLACIER, and is studied in a common R\&D program. The main difference, compared to the 
GLACIER case, is the large gain required in order to detect the few electrons which can be extracted following a nuclear recoil in LAr. In fact, 
the ionization from a nuclear recoil in liquid argon is highly quenched, and for the same energy loss a nuclear recoil produces about 4\% of 
the free ionization charge when compared to a beta electron. Considering that, with optimized low noise electronics, the energy threshold 
for the detection of an ionization signal (beta electron) without charge multiplication (gain=1) is $\sim$100 keV, it is clear that a gain of 
several hundreds is needed in order to detect a nuclear recoil of 30 keV. This large gain is in principle obtainable with a double stage LEM. 
The cathode high voltage of ArDM is provided using a 210 stages Greinacher (Cockroft-Walton) circuit, built to sustain up to 400 kV \cite
{Horikawa:2010bv}. It has been stably operated in liquid argon at 70 kV (0.6 kV/cm drift field in ArDM) in fully operational configuration.

\section{6 m$^3$} 

Liquid argon purity in a non-evacuable vessel is one of the critical design parameters for a GLACIER type detector. In the ICARUS 
approach, the TPC is housed in a vacuum vessel, and obtaining a good vacuum allows to remove air from the vessel prior to filling with LAr, 
help outgassing from the TPC materials and the vessel walls, and verify the integrity (tightness) of the vessel itself. 
Commercially available LAr, which is an acceptable starting point for farther purification in the liquid phase, is handled, transported and 
stored without using high vacuum vessels, and has an O$_2$ content of about 1 ppm. Therefore it looks very reasonable to expect that at 
least the removal of the air  can be easily achieved without vacuum, and air in a closed vessel can be efficiently displaced by a flow of Ar 
gas. 
This technique for reducing the air concentration down to the ppm level using a 6 m$^3$ vessel has been tested experimentally, and results 
are presented in detail in \cite{Curioni:2010gd}.
This work is the first step of a targeted experimental activity to verify the feasibility and to study the practical details of purging a large, fully 
instrumented, non-evacuable vessel from air down to ppm level air concentration through flushing with Ar gas, and to fill with ultra-pure 
liquid Ar (ppt level concentration of residual impurities).

\section{T32@J-PARC}

\begin{figure}[h]
\includegraphics[width=24pc]{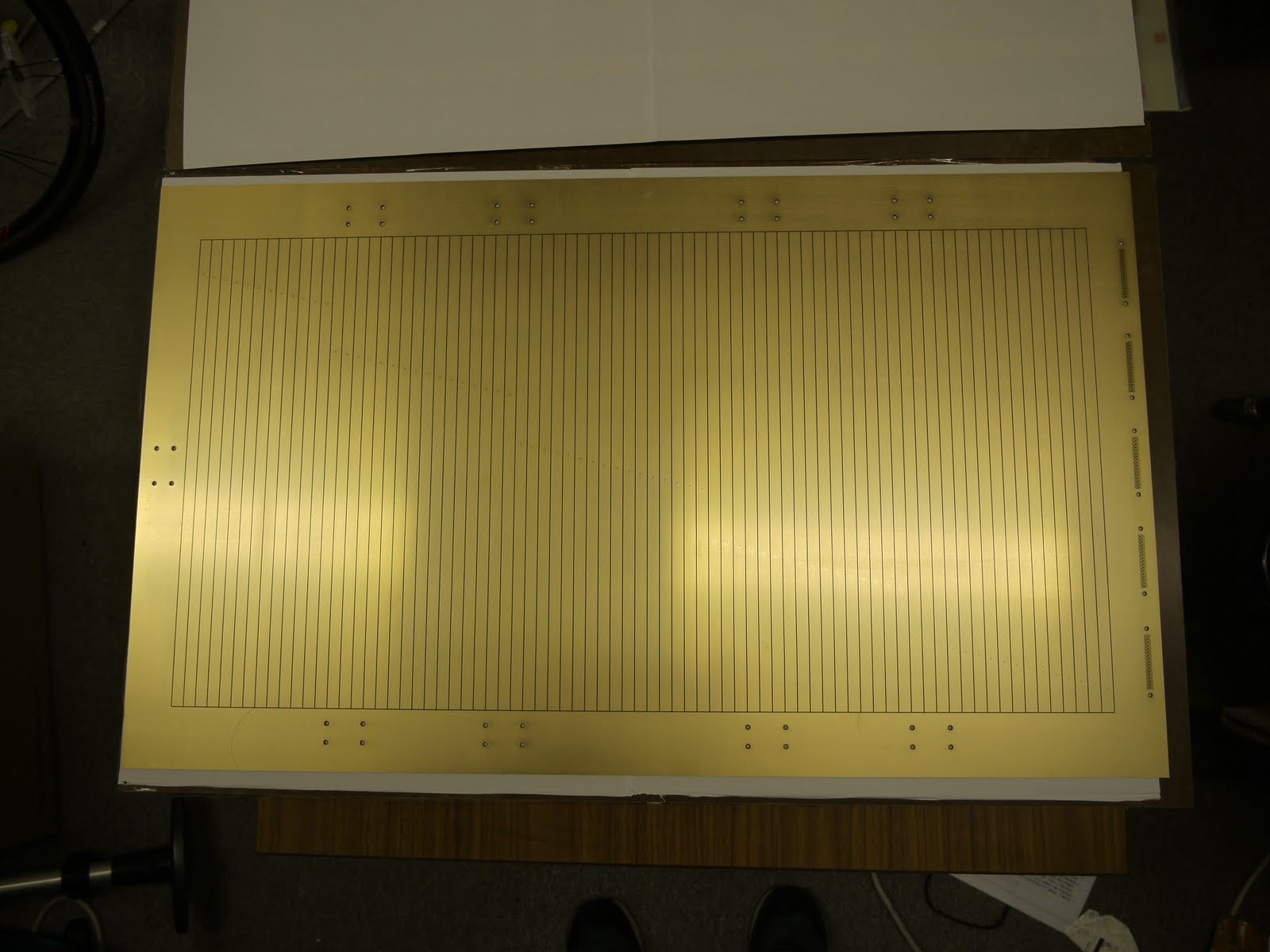}
\includegraphics[width=18pc]{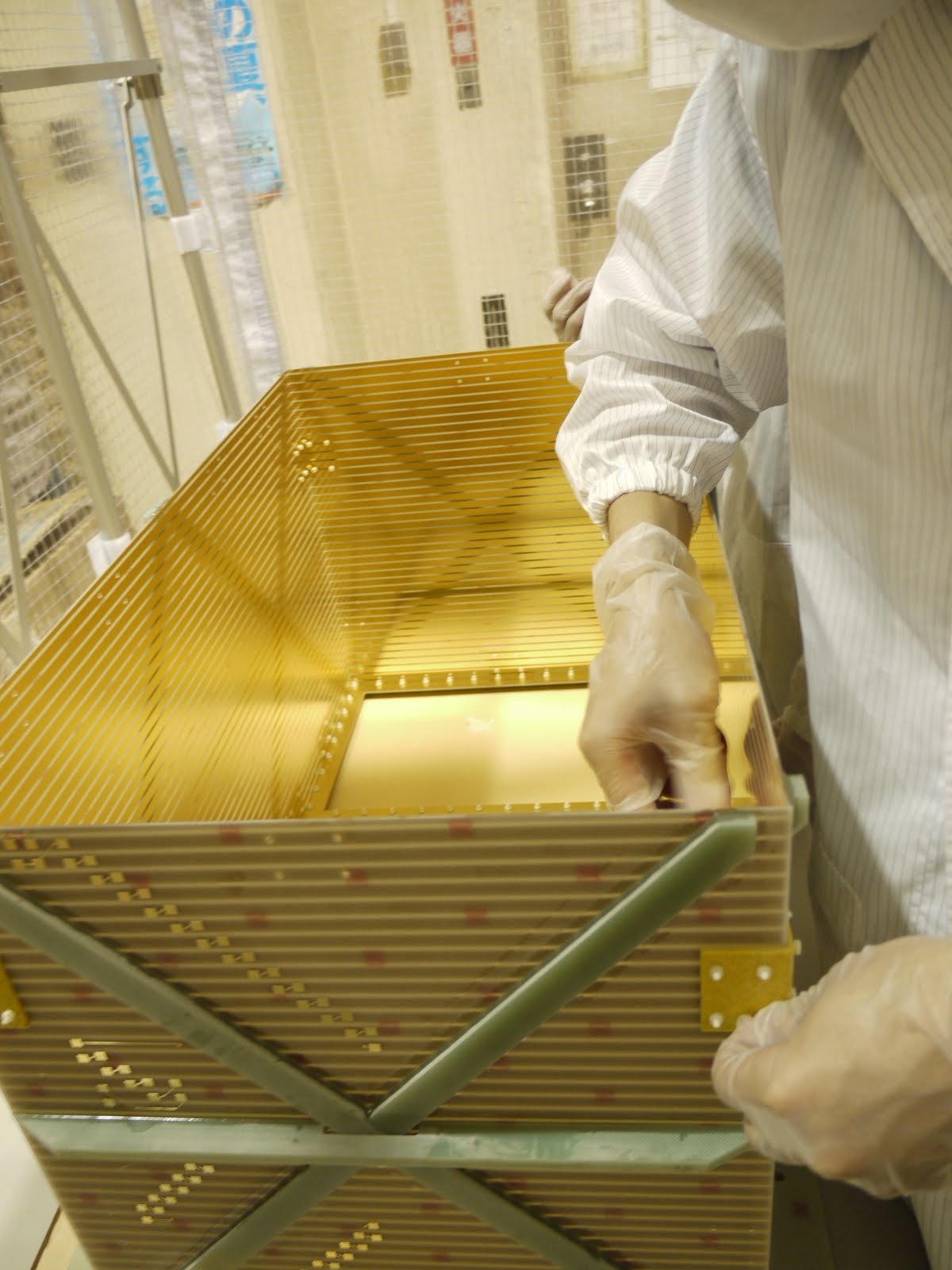}
\caption{\label{f_250L} Left: Readout plane of the T32@J-PARC LAr TPC, with 76 1 cm wide strips. Right: Field cage. }
\end{figure}

A 250L LAr TPC is being developed in a collaboration between KEK, Iwate University, Waseda University and ETH, for the T32@J-PARC 
experiment. The detector itself is an innovative step in the direction of a dual-phase, GLACIER-like LAr TPC. The goal of the experimental 
project is to perform measurements with well defined charged particle beams, in order to benchmark the performance of the detector and 
develop analysis and software tools for LAr TPC detectors, starting from large samples of positrons, pions, kaons and protons in LAr.  A 
preliminary version of the detector has already been exposed to the K1.1BR beam in October 2010 and about 170,000 events have been 
collected. For the initial test, a cryogenic vessel from the MEG experiment\footnote{http://meg.web.psi.ch/} has been used, which has now 
been replicated and replaced with a new one. This preliminary version is equipped with a 1D readout plane, with 76 strips with 1 cm pitch, 
running orthogonal to the beam, working as single phase LAr TPC (see Fig.~\ref{f_250L}). 
As discussed, the full LEM readout is presently being commissioned at CERN. 

\section{Conclusions and outlook}
Several LAr detectors have been developed at CERN and KEK to address the R\&D needs of GLACIER. In particular the 3L setup has been 
a workhorse for developing the LEM charge readout. We are now working on devices at the ton-scale, like ArDM and T32@J-PARC.
In the near future, we plan to fully instrument a 6 m$^3$ device (10 ton scale) as a double phase imaging detector. 
On a longer timescale, we are considering a 1 kton detector with a well defined physics program, as the final step in a series of prototypes 
leading to the 100 kton scale.


\end{document}